\documentclass[reprint,superscriptaddress, prl, amsmath,amssymb, aps]{revtex4-2}

\pdfoutput=1

\usepackage{graphicx}%
\usepackage{dcolumn}%
\usepackage{bm}%
\usepackage[colorlinks= true, linkcolor=blue, citecolor=blue, urlcolor=blue]{hyperref}%
\usepackage{lipsum}
\usepackage{mathtools}
\usepackage{xcolor}
\usepackage{multirow}

\graphicspath{{./figs/}{./}}

\begin{document}

\title{Phase transitions beyond criticality: extending Ising universal \\scaling functions to describe entire phases}%

\author{David Hathcock}
\affiliation{IBM T. J. Watson Research Center, Yorktown Heights, NY 10598}

\author{James P. Sethna}
\affiliation{Laboratory of Atomic and Solid State Physics,  Cornell University, Ithaca, New York 14853, USA}

\date{\today}%

\begin{abstract}

Universal scaling laws only apply asymptotically near critical phase transitions. We propose a general scheme, based on normal form theory of renormalization group flows, for incorporating corrections to scaling that quantitatively describe the entire neighboring phases. Expanding Onsager's exact solution of the 2D Ising model about the critical point, we identify a special coordinate with radius of convergence covering the entire physical temperature range, $0<T<\infty$. Without an exact solution, we demonstrate that using solely the critical singularity with low- and high-temperature expansions leads to  exponentially converging approximations across all temperatures for both the 2D and 3D Ising free energies and the 3D magnetization.
We discuss challenges and opportunities for future work.

\end{abstract}

\pacs{Valid PACS appear here}%

\maketitle

Since its introduction 50 years ago, the renormalization group (RG) has revolutionized our understanding of systems exhibiting emergent scale invariance. It has seen success in capturing universality and scaling laws in equilibrium phase transitions \cite{wilson1971renormalization, fisher1998renormalization, goldenfeld2018lectures}, bifurcations and the onset of chaos \cite{feigenbaum1979universal, hirsch1982intermittency, hu1982exact, raju2018reexamining, hathcock2021reaction}, disordered and glassy systems \cite{mckay1982spin, parisi2001renormalization,ledoussal2004functional, dupuis2021nonperturbative,dupuis2020bose}, and systems driven out of equilibrium \cite{kardar1986dynamic,toner1995long,toner1998flocks,yu2021inverse}.
Despite these achievements, the renormalization group only describes asymptotic behavior near a critical point; fluctuations, correlations, and other properties of the phases far from criticality are morphed by corrections to scaling. In this Letter, we show that smooth coordinate transformations can extend the universal scaling function predicted by RG analysis to {\em quantitatively describe the entire phase}.

Why is this significant? If the method presented here can be extended to the 3D Ising universality class, it could provide a systematic solution to two famous problems with `no small parameters': liquids and nuclear matter. The ideal gas and the zero-temperature crystal can be perturbatively corrected to describe the crystal and gas phases; can we describe liquids by perturbatively correcting the 3D Ising universal scaling at the liquid-gas critical point? The QCD phase diagram for massive quarks as a function of temperature and chemical potential also has an Ising critical point~\cite{stephanov2005qcd}; incorporating information about the theory for massless quarks, zero chemical potential, and zero field could guide us to a complete description of nuclear matter.

The key insight in our work comes from normal form theory, which classifies dynamical systems in terms of minimal nonlinear flows \cite{murdock2003normal}. The renormalization group defines parameter flows under coarse graining, with the fixed point of these flows representing a self-similar critical point. Normal form theory unifies renormalization group analyses for different physical models under a single framework: it predicts that near the critical point, smooth changes in coordinates (in temperature, field, etc.) can be used to systematically remove higher order nonlinearities in RG flows \cite{raju2019normal, sethna2023normal}. Usually, all nonlinear terms can be removed, leading to linear flows and the well known power-law scaling near criticality. In certain dimensions, however, resonances (2D Ising) or marginal variables (4D Ising, 2D XY model) lead to non-removable nonlinear terms, producing logarithms, exponentials, or more exotic scaling. The relevant variables in the normal form determine the universal critical scaling, with irrelevant variables adding singular corrections to scaling and analytic corrections arising from the normal form coordinate transformation. Incorporating these corrections to scaling will map out the entire phase diagram. 

We apply this idea to the zero-field 2D Ising model, where Onsager's exact solution~\cite{onsager1944crystal} for the free energy enables quantitative tests of our ability to capture phases far from the critical point. 2D Ising is clearly the easiest case: it has no singular corrections to scaling to the free energy or magnetization at zero field~\footnote{Corrections to scaling due to irrelevant perturbations in the 2D Ising model have integer exponents and are indistinguishable from analytic corrections. These are known as \emph{redundant} variables; there exists a suitable RG with the zero-field 2D Ising model as a fixed point \cite{raju2018reexamining}.} and  
it has known symmetries and analytic structure, namely the low-high temperature duality. Nonetheless, we show that our approach can be extended to the 3D Ising model, including singular corrections and non-trivial amplitude ratios without knowledge of symmetries of the model.

Our work reveals: (1) coordinate choice has a substantial impact on convergence of normal form analytic corrections. Leveraging a special dual-symmetric temperature variable in 2D, we find a free energy expansion that converges for all temperatures. (2) Fitting the universal critical point scaling form to data far from the critical point (low- and high-temperature expansions) leads to more robust, uniform exponential convergence, both in 2D and in 3D. 
By showing that the free energy scaling of zero-field Ising models can be extended to the entire phase, we unveil the surprising physical nature of convergence of normal form theory and lend hope that a general-purpose method will be possible.

To start, we compute the exact normal form coordinate transformation for the 2D Ising model and explore how convergence depends on the choice of temperature variable.
The general nonlinear RG flows for temperature $t = T-T_c$ and free energy $f$ in zero field are
${d  f}/{d \ell} = 2  f -  A t^2  + \mathcal{O}(t^3)$ and ${d  t}/{d \ell} =  t + \mathcal{O}(t^2)$. Normal form theory shows that the nonlinear terms are nonuniversal: except for the quadratic term  \footnote{In the case of the 2D Ising model the quadratic temperature term in the free energy $At^2$ is non-removable due to a resonance: an eigenvalue of the of the RG flow is an integral linear combination of other eigenvalues, $\lambda_0 = \sum_i m_i \lambda_i$, with $m_i \in \mathbb{N}$ \cite{murdock2003normal, raju2019normal}.
  Specifically, temperature $\lambda_t=1$ and the free energy $\lambda_f = 2 = 2 \lambda_t$ are resonant.}
in $df/d\ell$ they can all be removed by a changing to a nonlinear scaling variable $t\to t(\tilde t)$ (with inverse transformation, $\tilde t(t) = a_1  t + a_2  t^2 + \cdots$) and subtracting an analytic background $\tilde f(t) =  f(t) -f_a(t)$; $a_1$ is used to scale the quadratic coefficient $A$ to one.
These flow equations imply the well-known log singularity near the critical point, $\tilde f(\tilde t) = -\frac{1}{2} \tilde t^2 \log \tilde t^2$.
To obtain the free energy in terms of the physical temperature $t$ we simply insert the inverse normal form coordinate transformation, $\tilde t(t)$, and add back the analytic background,
\begin{equation}
\label{scalingForm}  
f(t) = -(1/2)\tilde t(t)^2 \log \tilde t(t)^2 + f_a(t). 
\end{equation}
This expression extends the validity of the logarithmic scaling form to describe a region neighboring the critical point. Can it be extended to the entire ordered and disordered phases?

\begin{figure}[!t]
\includegraphics[width=\linewidth]{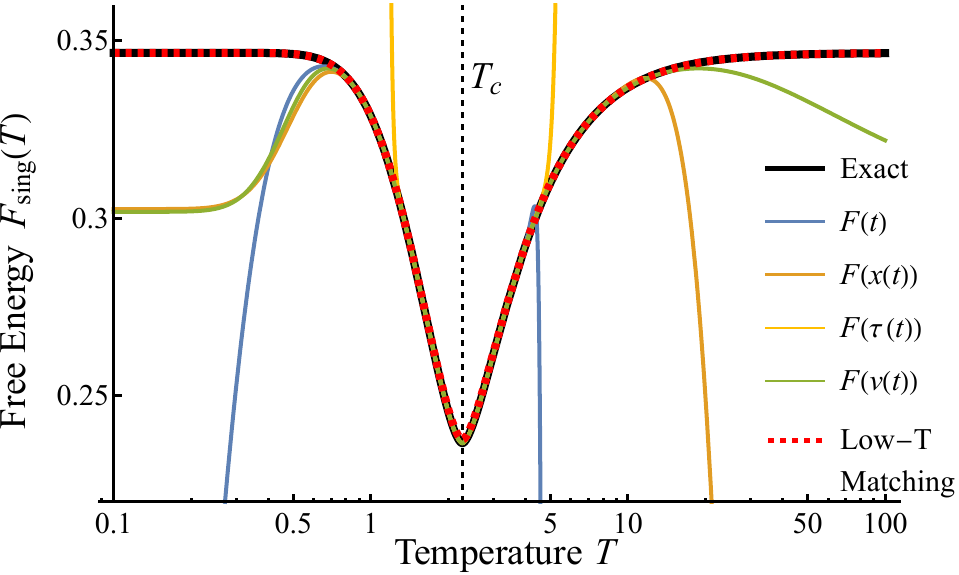} 
\caption{\label{coordinateExpansions} Capturing the entire phase with analytic corrections. Expanding Onsager's exact free energy (black line) in $t$, $x$, $\tau$, and $v$ (colored solid lines, $20^\text{th}$ order expansion) leads to variable radii of convergence. The convergence range for $\tau$ is limited by the the transformation $\tau(t)$, while the convergence in the other coordinates is set by the distance from the critical point to zero temperature, which covers the entire physical temperature range for the $v$-expansion. Determining the analytic corrections in $v$ by matching $a(t)$ and $b(t)$ in Eq.~\ref{abExpressions} to the low-temperature expansion accurately reproduces the free energy at all temperatures, even at low orders (red dashed line, $6^\text{th}$ order), see also Fig.~\ref{derivativesAndConvergence}.}
\end{figure}

To test this, we use Onsager's exact solution for the zero-field free energy of the 2D Ising model \cite{onsager1944crystal}, $-\beta F  = F_\text{sing}+ \frac{1}{2} \log (2 \cosh(2 \beta)^2)$, where
\begin{equation}\label{exactF}
\begin{split}
F_\text{sing} =
 \frac{1}{2 \pi} \int_0^\pi \log \left[ 1+\sqrt{1  -\frac{\cos^2(\theta)}{1+\tau^2}}  \right ]d\theta 
\quad  
\end{split}
\end{equation}
and $\tau = (1/\sinh 2 \beta - \sinh 2 \beta)/2$. This expression is known to have the form $a(t) \log t^2 +b(t)$, for some analytic functions $a(t)$ and $b(t)$. Comparing to the normal form prediction for the free energy we have
\begin{equation}\label{abExpressions}
a(t) = - \frac{1}{2} \tilde t(t)^2, \quad \,
b(t) = \frac{1}{2} \tilde t(t)^2 \log( \tilde t(t)^2/t^2) +f_a(t).
\end{equation}
The term $\log(\tilde t(t)^2/t^2)$ is analytic because $\tilde t(t) \approx a_1 t$ to leading order.
To explore the ability of our coordinate change to capture the high and low temperature phases, we expand the exact free energy around the critical point and deduce the analytic functions $\tilde t$ and $f_a$ (or equivalently $a$ and $b$). These expansions also allow us to infer the exact nonlinear RG flows for the 2D Ising model to arbitrary order~\cite{SM}.

Unlike an ordinary Taylor expansion, the normal form expansion about the critical singularity does not have radius of convergence simply determined by the distance to the nearest singularity in the complex plane. One might hope that the convergence is determined by physical boundaries: the distance to zero temperature, infinity temperature, or another critical point, for example. We have found that the choice of coordinate in computing the expansions of Eq.~(\ref{scalingForm}) has a substantial impact on the radius of convergence (see Fig.~\ref{coordinateExpansions}). Indeed, in most of the cases discussed below, the critical point is closest to zero temperature, with this distance setting the radius of convergence. While identifying a good normal form coordinate is useful, we will show later that it is non-essential: we introduce an alternative scheme for incorporating corrections to scaling with exponential convergence even for coordinates with limited radius of convergence.

\begin{figure*}[t]
\includegraphics[width=\linewidth]{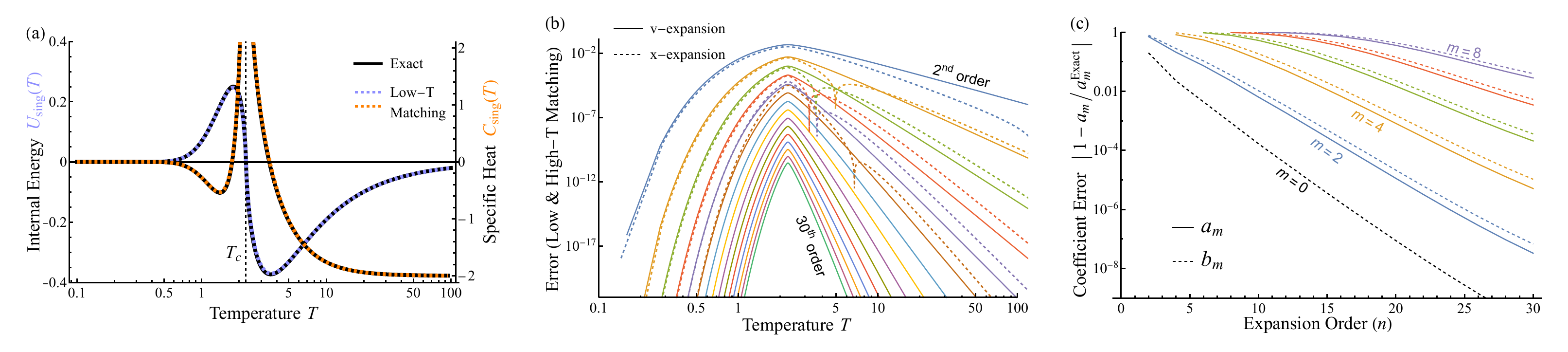}
\caption{\label{derivativesAndConvergence} Free energy derivatives and exponential convergence. (a) Approximating the free energy by matching to low-temperature expansions accurately captures the internal energy $U_\text{sing} = \partial F_\text{sing}/\partial \beta$ and the specific heat $C_\text{sing} = \partial^2 F_\text{sing}/\partial \beta^2$ across all temperatures, including the logarithmic singularity in the latter. (b) The approximation error $|F_\text{approx} - F_\text{sing}|$ for the low-temperature matching expansion in $v$ (even orders $n=2-30$, solid lines) or the low- and high-temperature matching expansions in $x$ (odd orders $n=3-13$, dashed lines). Adding terms leads to exponential convergence to the exact free energy. (c) The expansion coefficients $a_m$ (solid) and $b_m$ (dashed), Eq.~(\ref{expansionSeries}), also converge exponentially to their exact values (shown for the $v$-expansion). If more information at the critical point is known (e.g. low order coefficients from Onsager's solution), a hybrid matching scheme leads to accelerated convergence. {\color{black} Conversely, if $T_c$ is not known, it can be fit with only slightly slower convergence~\cite{SM}.}}
\end{figure*}

Expanding $\tilde t$ and $f_a$ in temperature $t=T-T_c$, for example, converges in an estimated range $0<T<2 T_c$~\cite{SM}. As anticipated above, the distance to zero temperature sets the radius of convergence. Another natural choice is the low-temperature coordinate $X = \exp(-2/T)$, used for cluster expansions of Ising models in all dimensions. Expanding in $x=X-X_c$, we find a larger convergence range $0<X<2 X_c$ (corresponding to $0<T\lesssim 4.7 T_c$), again set by the distance to zero temperature. 

These results prompt the question: can we chose a coordinate for which the expansion of the normal form expansion converges for all temperatures? The solution is to choose a coordinate that respects the low-high temperature duality of the 2D Ising model, with the critical point lying halfway between zero and infinite temperatures.

One such coordinate is $\tau$ that appears in Eq.~(\ref{exactF}),
which is also a natural choice for expansions near the critical point \cite{orrick2001critical,caselle2002irrelevant,chan2011ising}. Unfortunately, as shown in Fig.~\ref{coordinateExpansions} (yellow line), the convergence in $\tau$ is actually worse than the coordinate choices mentioned above. Instead of the convergence being set by the distance to zero temperature, it is restricted by the radius of convergence of the mapping $T \to T(\tau) = -2/\sinh^{-1}(\tau - \sqrt{1+ \tau^2})$, leading to a convergence range $-1<\tau <1$ ($0.55 T_c \lesssim T \lesssim 2.2 T_c$).

To identify a more suitable coordinate that respects the 2D Ising duality, we look in the complex temperature plane. Here the partition function has zeros (sometimes called Fisher zeros \cite{fisher1964nature, dolan2001thin,denbleyker2010fisher}), which give rise to a branch cut in the free energy and poles in the internal energy and specific heat. For the 2D Ising model it is well known that these zeros 
form two circles in the complex $X$-plane, one of which: $X = -1 + \sqrt{2} e^{i \theta}$ ($0\leq \theta \leq 2 \pi$) intersects the critical point, $X_c = -1 +\sqrt{2}$~\cite{SM}. %
This analytic structure motivates a new coordinate that unwraps the circle of Fisher zeros, so that they lie on a straight line in the complex plane. We define $V$ using the linear fractional transformation that carries the circle to a vertical line intersecting $V_c = X_c$,
\begin{equation}\label{Vcoordinate}
V = \frac{5 - 3 \sqrt{2} +X}{1+ \sqrt{2}+ X}.
\end{equation}
Our new coordinate extends the self-dual symmetry of the 2D Ising model to the complex plane: the singular part of the free energy, Eq.~(\ref{exactF}) is symmetric under the transformation $v \to -v$ ($v=V-V_c$) that maps between dual low and high temperatures. 
As shown in Fig.~\ref{coordinateExpansions}, the expansion of the free energy in $v$ has the largest radius of convergence, with the scaling of the coefficients indicating that it converges for all physical temperatures~\cite{SM}.

In practice, even with a radius of convergence covering all temperatures, we may still fail to capture the behavior at very low or high temperatures unless we can expand to arbitrary orders. Furthermore, for most models and experimental systems we have limited information at the critical point, e.g. only the asymptotic scaling form and critical exponents. Can we produce an approximation that is uniformly valid across all temperatures and requires minimal knowledge of the critical scaling?

To resolve these challenges, we take another approach: rather than determining $\tilde t$ and $f_a$ in Eq.~(\ref{scalingForm}) by expanding at the critical point, we fit the coefficients by matching to low-temperature cluster expansions of the free energy. Such expansions are comparatively easy to compute for Ising models in all dimensions. This approach has some similarity to previous extended scaling analyses~\cite{campbell2007extended,campbell2008extended,campbell2008extendedZeroTemp,campbell2011extended}, but integrates normal form scaling functions and is easily extended to high orders to describe entire phases. Here, we again write the free energy as $f(v) = a(v) \log v^2 + b(v)$ [via Eq.~(\ref{abExpressions})] %
and expand the coefficients $a$ and $b$, 
\begin{equation}\label{expansionSeries}
    a(v) = \sum_{n=1}^{\infty} a_{n} v^{2n} \quad \quad b(v) = \sum_{n=0}^{\infty} b_{n} v^{2n}
\end{equation}
by matching derivatives at zero temperature. With no higher-order terms inside the log, this amounts to solving a linear system.

Our low-temperature matching approach only requires knowledge of the asymptotic scaling form at the critical point, with most of the information coming from deep within the low-temperature phase. Since the correct low-temperature behavior is guaranteed (as is the high-temperature behavior if we use the coordinate $v$ that respects the 2D Ising duality) and the expansion has the correct logarithmic singularity at the critical point, we see uniform convergence across all temperatures. By sixth-order (Fig.~\ref{coordinateExpansions}, red dashed line), the approximation differs from the true free energy by at most $0.5\%$ and accurately captures the internal energy and specific heat, including the logarithmic divergence in the latter (Fig.~\ref{derivativesAndConvergence}a). Remarkably, as we add terms to the expansion, we see exponential convergence to the exact solution (Fig.~\ref{derivativesAndConvergence}b, solid lines). In contrast to expanding at the critical point (Fig.~\ref{coordinateExpansions}), this approach varies the higher-order expansion coefficients to fit the behavior at zero temperature. As we add terms, however, these coefficients still converge exponentially to those obtained by expanding the exact solution (Fig.~\ref{derivativesAndConvergence}c). 

While it is helpful to use the expansion coordinate $v$, leveraging our knowledge of the complex analytic structure and the duality of the 2D Ising model, this is not necessary. If instead we expand in the coordinate $x$ and match to both low- and high- temperature cluster expansions, we obtain similar exponential convergence across all temperatures (Fig.~\ref{derivativesAndConvergence}b, dashed lines). The expansion in $x$ requires matching twice as many terms to achieve comparable accuracy to the $v$-expansion and the convergence appears to slow slightly at high orders. Nonetheless, its success indicates the approach will generalize to unsolvable models and experimental systems, where we may not have knowledge of analytic structure and symmetries. Fitting the expansion coefficients to data {\em outside the radius of convergence of the series generated at the critical point} extends the range of exponential convergence.

Finally, we apply this approach to the 3D Ising model, which has two major complexities beyond analytic corrections: a non-trivial amplitude ratio and singular corrections. The 3D free energy and magnetization are approximately \cite{SM},
\begin{equation}\label{3DFreeEnergyMagnetization}
\begin{split}
    f(t) &\approx  A_\pm  |\tilde t(t)|^{2-\alpha} +  B_\pm  \tilde u(t) |\tilde t(t)|^{2-\alpha + \theta} + f_a(t),\\
    m(t) &\approx \tilde h_1(t)(|\tilde t(t)|^\beta + C \tilde u(t) |\tilde t(t)|^{\beta + \theta}) \quad \quad t<0,
\end{split}
\end{equation}
where $\alpha \approx 0.11008$, 
$\beta \approx 0.326419$, and $\theta = \nu \omega \approx 0.52266$ are the 3D critical exponents for the specific heat, 
magnetization and leading irrelevant perturbation $u$ \cite{kos2016precision, komargodski2017random}. The nonlinear scaling variable for the irrelevant perturbation is denoted $\tilde u(t) \approx u_0 + u_1 t + \cdots$, while $f_a(t)$ is the analytic background. The constants $A_\pm$ and $B_\pm$ depend on the sign of $t = T-T_c$ and have universal ratios, with the former known to a few digits $A_+/A_- \approx 0.536\pm 0.002$ \cite{hasenbusch2010universal}. 
We neglect terms higher order in $u$ and contributions from sub-leading singular corrections. Here we provide early results indicating that both amplitude ratios and singular corrections can be incorporated into our normal-form expansion scheme.

Working in the low-temperature coordinate $x$, we approximate the free energy and magnetization by fitting the expansions $\tilde t$, $\tilde u$, $\tilde h_1$ and $f_a$ to match to high- and low- temperature expansions \cite{guttmann1993series, guttmann1994high, bhanot1994series}. For the free energy, we incorporate the universal amplitude ratio, but neglect the correction to scaling; for the magnetization we use a constant-coefficient singular correction $\tilde u(x) \approx u_0$. To improve convergence, we also fit the well-known complex-temperature pole $x_p<0$ with $|x_p|<x_c$ (see Supplemental Material for fitting details~\cite{SM}). A comprehensive study of fitting procedure and the impact of relative orders of expansion will be an interesting direction for future work.

\begin{figure}[t]
\includegraphics[width=\linewidth]{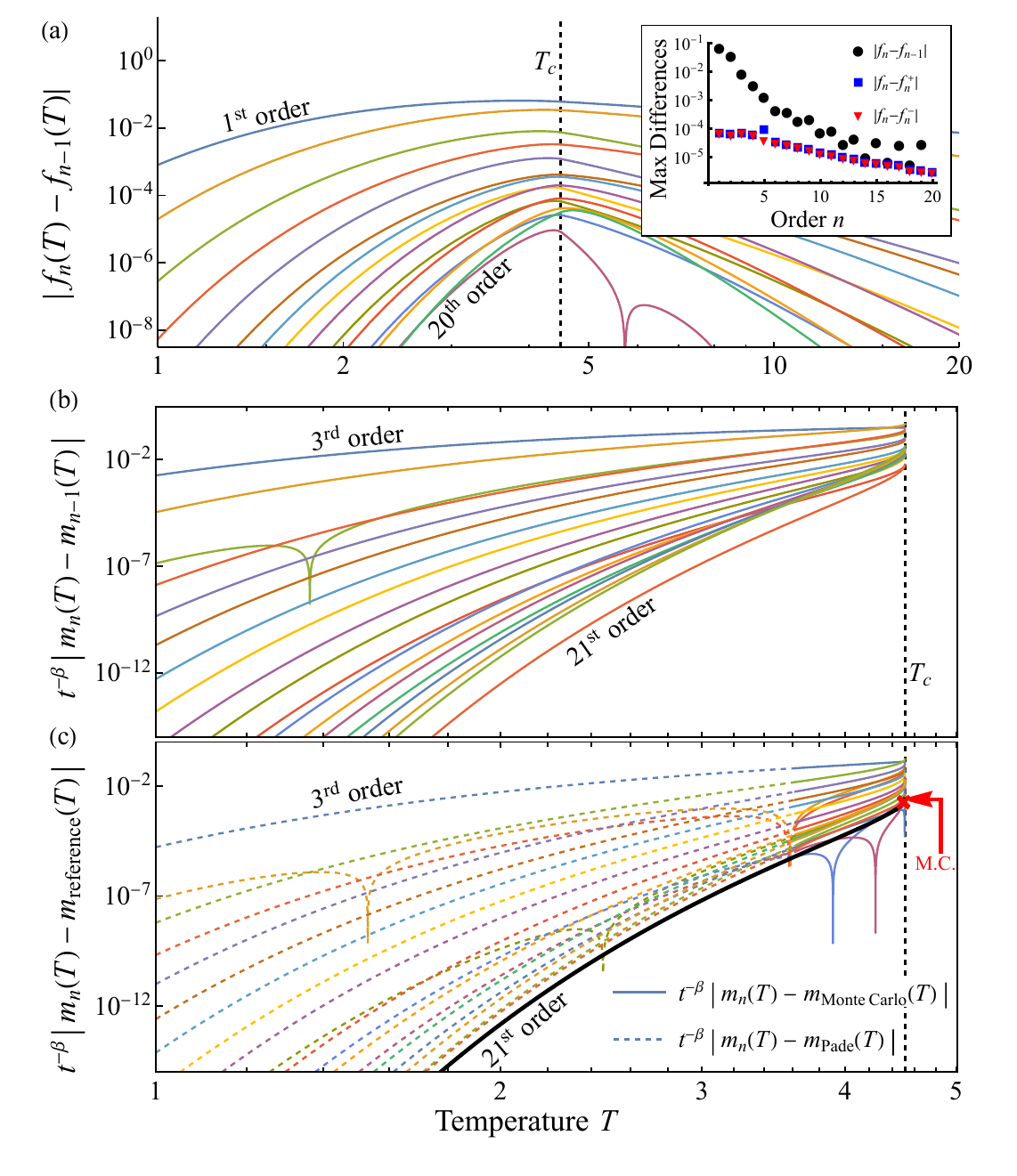}
\caption{\label{3DIsingApproximations} Approximations of the 3D Ising free energy and magnetization, obtained by expanding Eq.~(\ref{3DFreeEnergyMagnetization}) and matching to cluster expansions. (a) The difference between subsequent approximations of the free energy. The inset shows the maximum difference $|f_n-f_{n-1}|$ and the maximum distances to approximations $f^{\pm}_n(T)$ using perturbed amplitude ratios $A_+/A_- = 0.536 \pm 0.002$.  (b) The difference between subsequent approximations of the magnetization. (c) Comparison of the magnetization expansions to a (9,9)-Pad\'e approximation~\cite{essam1963pade} (dashed) and fits to Monte Carlo simulations~\cite{talapov1996magnetization} (solid). The later includes the leading singular correction and is more accurate for $T_c>T\gtrsim 3.6$. The thick black line shows the best converged difference from (b), which we also compare to the high-precision Monte Carlo estimates of the magnetization near $T_c$~\cite{hasenbusch2012thermodynamic} (red cross, expanded plot in SM~\cite{SM}). In (b-c) we scale by the leading power-law $t^\beta = (T_c-T)^\beta$.}
\end{figure}

Fig.~\ref{3DIsingApproximations} shows the difference between subsequent approximations, which converge exponentially. The free energy converges up to the limited precision of the amplitude ratio, while the magnetization shows better precision than the traditional d-log Pad\'e approach near the critical point~\cite{essam1963pade}, where it agrees excellently with high-precision Monte Carlo studies~\cite{talapov1996magnetization,hasenbusch2012thermodynamic, SM}. The approximations are naturally least accurate near the critical point because we are fitting information at zero temperature and where the behavior is most sensitive to the precision of the critical exponents, amplitude ratio and $T_c$. {\color{black} While the approximations are best if they incorporate high-precision information about the critical point (critical temperature, universal exponents, amplitude ratios, etc.), the normal form transformation can be adapted to incorporate these quantities as fitting parameters~\cite{SM}.}

Looking forward, an even more complicated assortment of singular corrections appear in the 2D Ising susceptibility ~\cite{orrick2001critical, chan2011ising}, 
\begin{equation}
    \chi = \frac{A_\pm(t)} {\tilde t(t)^{7/4}} + \sum_{q=0}^\infty \sum_{p=0}^{\lfloor \sqrt{q} \rfloor} b_{q,p} t^q (\log |t|)^p,
\end{equation}
where $A_\pm(t)$ is an analytic function (that depends on whether $T>T_c$ or $T<T_c$), but includes a variety of integer-exponent singular corrections that can be directly associated with irrelevant fields~\cite{caselle2002irrelevant, chan2011ising}. The second term includes logarithmic singularities that correct the leading power law. The higher powers of logs may arise due to resonances involving irrelevant parameters, similar to that between the free energy and temperature, but the precise form of these resonances remains unknown \cite{raju2018aspects, clement2019respect}. Given the expansions of the susceptibility at low- and high-temperatures to over 2000 orders \cite{chan2011ising}, it will provide further testing ground for our method of low- and high-temperature matching in future studies. For example, should we include many logarithmic corrections in our expansion, or are the high-order logs suitably approximated by analytic corrections? Indeed near the critical point they are subdominant and far from the critical point they are analytic.

Understanding the susceptibility is the first step in mapping out the full field-dependent 2D Ising free energy, $\tilde f(t,h) = \tilde t^{2-\alpha} \mathcal{F}(\tilde h/\tilde t^{y_h}) -\frac{1}{2}\tilde t^2 \log \tilde t^2$. To this end, joint low-temperature high-field and high-temperature low-field expansions are well established \cite{sykes1965derivation, katsura1977high} and an exponentially convergent approximation for the scaling function $\mathcal{F}$ has recently been developed \cite{kentDobias2017smooth, sethna2023normal}. %

Beyond Ising models, this approach should apply to a broad class of theoretical and experimental systems where the critical exponents and universal scaling forms have been determined elsewhere. Fitting to measurements, simulations, or perturbative calculations far from the critical point may enable extension of critical scaling forms to the entire phase diagram. Potential target applications for this approach include the liquid-gas transition and lattice QCD (3D Ising \cite{stephanov2005qcd, borsanyi2021lattice}), cell membranes (2D Ising \cite{machta2011minimal, machta2012critical}), and superconductivity \cite{shankar1994renormalization, abanov2001spin}, as well as non-equilibrium systems in the Kadar-Parisi-Zhang (KPZ) \cite{kardar1986dynamic} or flocking \cite{toner1995long, toner1998flocks} universality classes. 

{\color{black}Our results are directly applicable to systems with a single critical transition, but we are optimistic that the theory can be generalized to map out entire phases in systems with multiple transitions. In particular, for each critical point one can write down the (possibly nonlinear) normal form renormalization group flows. Combining these yields a minimal nonlinear form that captures the critical scaling at each transition~\cite{sethna2023normal}. A single normal form coordinate change (which might be determined using data or calculations deep within each phase) applies across all transitions to extend scaling functions beyond criticality. This approach might be especially suited for nuclear matter: lattice QCD methods are powerful for massless quarks, revealing first- and second-order transitions in the 3D tricritical Ising universality class~\cite{stephanov2005qcd}. A normal-form theory of the crossover to the 3D Ising QCD critical point, plus the analytic corrections discussed here matched to zero baryo-chemical potential computations~\cite{ borsanyi2021lattice}, could generate quantitative predictions for the whole strong-interaction phase diagram.}
\newline

\begin{acknowledgments}
Thanks to Peter Lepage, Jacques Perk, Archishman Raju, Jaron Kent-Dobias, and Stephen Thornton for helpful discussions and comments on the manuscript. D.H. was partially supported by an NSF Graduate Research Fellowship Grant No. DGE-2139899. J.P.S. was supported by NSF DMR-2327094.
\end{acknowledgments}
\bibliography{references}

\begin{thebibliography}{54}%
\makeatletter
\providecommand \@ifxundefined [1]{%
 \@ifx{#1\undefined}
}%
\providecommand \@ifnum [1]{%
 \ifnum #1\expandafter \@firstoftwo
 \else \expandafter \@secondoftwo
 \fi
}%
\providecommand \@ifx [1]{%
 \ifx #1\expandafter \@firstoftwo
 \else \expandafter \@secondoftwo
 \fi
}%
\providecommand \natexlab [1]{#1}%
\providecommand \enquote  [1]{``#1''}%
\providecommand \bibnamefont  [1]{#1}%
\providecommand \bibfnamefont [1]{#1}%
\providecommand \citenamefont [1]{#1}%
\providecommand \href@noop [0]{\@secondoftwo}%
\providecommand \href [0]{\begingroup \@sanitize@url \@href}%
\providecommand \@href[1]{\@@startlink{#1}\@@href}%
\providecommand \@@href[1]{\endgroup#1\@@endlink}%
\providecommand \@sanitize@url [0]{\catcode `\\12\catcode `\$12\catcode
  `\&12\catcode `\#12\catcode `\^12\catcode `\_12\catcode `\%12\relax}%
\providecommand \@@startlink[1]{}%
\providecommand \@@endlink[0]{}%
\providecommand \url  [0]{\begingroup\@sanitize@url \@url }%
\providecommand \@url [1]{\endgroup\@href {#1}{\urlprefix }}%
\providecommand \urlprefix  [0]{URL }%
\providecommand \Eprint [0]{\href }%
\providecommand \doibase [0]{https://doi.org/}%
\providecommand \selectlanguage [0]{\@gobble}%
\providecommand \bibinfo  [0]{\@secondoftwo}%
\providecommand \bibfield  [0]{\@secondoftwo}%
\providecommand \translation [1]{[#1]}%
\providecommand \BibitemOpen [0]{}%
\providecommand \bibitemStop [0]{}%
\providecommand \bibitemNoStop [0]{.\EOS\space}%
\providecommand \EOS [0]{\spacefactor3000\relax}%
\providecommand \BibitemShut  [1]{\csname bibitem#1\endcsname}%
\let\auto@bib@innerbib\@empty
\bibitem [{\citenamefont {Wilson}(1971)}]{wilson1971renormalization}%
  \BibitemOpen
  \bibfield  {author} {\bibinfo {author} {\bibfnamefont {K.~G.}\ \bibnamefont
  {Wilson}},\ }\bibfield  {title} {\bibinfo {title} {Renormalization group and
  critical phenomena. {I. R}enormalization group and the {Kadanoff} scaling
  picture},\ }\href {https://doi.org/10.1103/PhysRevB.4.3174} {\bibfield
  {journal} {\bibinfo  {journal} {Phys. Rev. B}\ }\textbf {\bibinfo {volume}
  {4}},\ \bibinfo {pages} {3174} (\bibinfo {year} {1971})}\BibitemShut
  {NoStop}%
\bibitem [{\citenamefont {Fisher}(1998)}]{fisher1998renormalization}%
  \BibitemOpen
  \bibfield  {author} {\bibinfo {author} {\bibfnamefont {M.~E.}\ \bibnamefont
  {Fisher}},\ }\bibfield  {title} {\bibinfo {title} {Renormalization group
  theory: Its basis and formulation in statistical physics},\ }\href
  {https://doi.org/10.1103/RevModPhys.70.653} {\bibfield  {journal} {\bibinfo
  {journal} {Rev. Mod. Phys.}\ }\textbf {\bibinfo {volume} {70}},\ \bibinfo
  {pages} {653} (\bibinfo {year} {1998})}\BibitemShut {NoStop}%
\bibitem [{\citenamefont {Goldenfeld}(2018)}]{goldenfeld2018lectures}%
  \BibitemOpen
  \bibfield  {author} {\bibinfo {author} {\bibfnamefont {N.}~\bibnamefont
  {Goldenfeld}},\ }\href@noop {} {\emph {\bibinfo {title} {Lectures on phase
  transitions and the renormalization group}}}\ (\bibinfo  {publisher} {CRC
  Press},\ \bibinfo {year} {2018})\BibitemShut {NoStop}%
\bibitem [{\citenamefont {Feigenbaum}(1979)}]{feigenbaum1979universal}%
  \BibitemOpen
  \bibfield  {author} {\bibinfo {author} {\bibfnamefont {M.~J.}\ \bibnamefont
  {Feigenbaum}},\ }\bibfield  {title} {\bibinfo {title} {The universal metric
  properties of nonlinear transformations},\ }\href
  {https://doi.org/10.1007/BF01107909} {\bibfield  {journal} {\bibinfo
  {journal} {Journal of Statistical Physics}\ }\textbf {\bibinfo {volume}
  {21}},\ \bibinfo {pages} {669} (\bibinfo {year} {1979})}\BibitemShut
  {NoStop}%
\bibitem [{\citenamefont {Hirsch}\ \emph {et~al.}(1982)\citenamefont {Hirsch},
  \citenamefont {Nauenberg},\ and\ \citenamefont
  {Scalapino}}]{hirsch1982intermittency}%
  \BibitemOpen
  \bibfield  {author} {\bibinfo {author} {\bibfnamefont {J.}~\bibnamefont
  {Hirsch}}, \bibinfo {author} {\bibfnamefont {M.}~\bibnamefont {Nauenberg}},\
  and\ \bibinfo {author} {\bibfnamefont {D.}~\bibnamefont {Scalapino}},\
  }\bibfield  {title} {\bibinfo {title} {Intermittency in the presence of
  noise: A renormalization group formulation},\ }\href
  {https://doi.org/https://doi.org/10.1016/0375-9601(82)90165-7} {\bibfield
  {journal} {\bibinfo  {journal} {Physics Letters A}\ }\textbf {\bibinfo
  {volume} {87}},\ \bibinfo {pages} {391} (\bibinfo {year} {1982})}\BibitemShut
  {NoStop}%
\bibitem [{\citenamefont {Hu}\ and\ \citenamefont
  {Rudnick}(1982)}]{hu1982exact}%
  \BibitemOpen
  \bibfield  {author} {\bibinfo {author} {\bibfnamefont {B.}~\bibnamefont
  {Hu}}\ and\ \bibinfo {author} {\bibfnamefont {J.}~\bibnamefont {Rudnick}},\
  }\bibfield  {title} {\bibinfo {title} {Exact solutions to the {F}eigenbaum
  renormalization-group equations for intermittency},\ }\href
  {https://doi.org/10.1103/PhysRevLett.48.1645} {\bibfield  {journal} {\bibinfo
   {journal} {Phys. Rev. Lett.}\ }\textbf {\bibinfo {volume} {48}},\ \bibinfo
  {pages} {1645} (\bibinfo {year} {1982})}\BibitemShut {NoStop}%
\bibitem [{\citenamefont {Raju}\ and\ \citenamefont
  {Sethna}(2018)}]{raju2018reexamining}%
  \BibitemOpen
  \bibfield  {author} {\bibinfo {author} {\bibfnamefont {A.}~\bibnamefont
  {Raju}}\ and\ \bibinfo {author} {\bibfnamefont {J.~P.}\ \bibnamefont
  {Sethna}},\ }\bibfield  {title} {\bibinfo {title} {Reexamining the
  renormalization group: Period doubling onset of chaos},\ }\href@noop {}
  {\bibfield  {journal} {\bibinfo  {journal} {arXiv preprint arXiv:1807.09517}\
  } (\bibinfo {year} {2018})}\BibitemShut {NoStop}%
\bibitem [{\citenamefont {Hathcock}\ and\ \citenamefont
  {Sethna}(2021)}]{hathcock2021reaction}%
  \BibitemOpen
  \bibfield  {author} {\bibinfo {author} {\bibfnamefont {D.}~\bibnamefont
  {Hathcock}}\ and\ \bibinfo {author} {\bibfnamefont {J.~P.}\ \bibnamefont
  {Sethna}},\ }\bibfield  {title} {\bibinfo {title} {Reaction rates and the
  noisy saddle-node bifurcation: Renormalization group for barrier crossing},\
  }\href {https://doi.org/10.1103/PhysRevResearch.3.013156} {\bibfield
  {journal} {\bibinfo  {journal} {Phys. Rev. Res.}\ }\textbf {\bibinfo {volume}
  {3}},\ \bibinfo {pages} {013156} (\bibinfo {year} {2021})}\BibitemShut
  {NoStop}%
\bibitem [{\citenamefont {McKay}\ \emph {et~al.}(1982)\citenamefont {McKay},
  \citenamefont {Berker},\ and\ \citenamefont {Kirkpatrick}}]{mckay1982spin}%
  \BibitemOpen
  \bibfield  {author} {\bibinfo {author} {\bibfnamefont {S.~R.}\ \bibnamefont
  {McKay}}, \bibinfo {author} {\bibfnamefont {A.~N.}\ \bibnamefont {Berker}},\
  and\ \bibinfo {author} {\bibfnamefont {S.}~\bibnamefont {Kirkpatrick}},\
  }\bibfield  {title} {\bibinfo {title} {Spin-glass behavior in frustrated
  {I}sing models with chaotic renormalization-group trajectories},\ }\href
  {https://doi.org/10.1103/PhysRevLett.48.767} {\bibfield  {journal} {\bibinfo
  {journal} {Phys. Rev. Lett.}\ }\textbf {\bibinfo {volume} {48}},\ \bibinfo
  {pages} {767} (\bibinfo {year} {1982})}\BibitemShut {NoStop}%
\bibitem [{\citenamefont {Parisi}\ \emph {et~al.}(2001)\citenamefont {Parisi},
  \citenamefont {Petronzio},\ and\ \citenamefont
  {Rosati}}]{parisi2001renormalization}%
  \BibitemOpen
  \bibfield  {author} {\bibinfo {author} {\bibfnamefont {G.}~\bibnamefont
  {Parisi}}, \bibinfo {author} {\bibfnamefont {R.}~\bibnamefont {Petronzio}},\
  and\ \bibinfo {author} {\bibfnamefont {F.}~\bibnamefont {Rosati}},\
  }\bibfield  {title} {\bibinfo {title} {Renormalization group approach to spin
  glass systems},\ }\href {https://doi.org/10.1007/s100510170171} {\bibfield
  {journal} {\bibinfo  {journal} {The European Physical Journal B - Condensed
  Matter and Complex Systems}\ }\textbf {\bibinfo {volume} {21}},\ \bibinfo
  {pages} {605} (\bibinfo {year} {2001})}\BibitemShut {NoStop}%
\bibitem [{\citenamefont {Le~Doussal}\ \emph {et~al.}(2004)\citenamefont
  {Le~Doussal}, \citenamefont {Wiese},\ and\ \citenamefont
  {Chauve}}]{ledoussal2004functional}%
  \BibitemOpen
  \bibfield  {author} {\bibinfo {author} {\bibfnamefont {P.}~\bibnamefont
  {Le~Doussal}}, \bibinfo {author} {\bibfnamefont {K.~J.}\ \bibnamefont
  {Wiese}},\ and\ \bibinfo {author} {\bibfnamefont {P.}~\bibnamefont
  {Chauve}},\ }\bibfield  {title} {\bibinfo {title} {Functional renormalization
  group and the field theory of disordered elastic systems},\ }\href
  {https://doi.org/10.1103/PhysRevE.69.026112} {\bibfield  {journal} {\bibinfo
  {journal} {Phys. Rev. E}\ }\textbf {\bibinfo {volume} {69}},\ \bibinfo
  {pages} {026112} (\bibinfo {year} {2004})}\BibitemShut {NoStop}%
\bibitem [{\citenamefont {Dupuis}\ \emph {et~al.}(2021)\citenamefont {Dupuis},
  \citenamefont {Canet}, \citenamefont {Eichhorn}, \citenamefont {Metzner},
  \citenamefont {Pawlowski}, \citenamefont {Tissier},\ and\ \citenamefont
  {Wschebor}}]{dupuis2021nonperturbative}%
  \BibitemOpen
  \bibfield  {author} {\bibinfo {author} {\bibfnamefont {N.}~\bibnamefont
  {Dupuis}}, \bibinfo {author} {\bibfnamefont {L.}~\bibnamefont {Canet}},
  \bibinfo {author} {\bibfnamefont {A.}~\bibnamefont {Eichhorn}}, \bibinfo
  {author} {\bibfnamefont {W.}~\bibnamefont {Metzner}}, \bibinfo {author}
  {\bibfnamefont {J.}~\bibnamefont {Pawlowski}}, \bibinfo {author}
  {\bibfnamefont {M.}~\bibnamefont {Tissier}},\ and\ \bibinfo {author}
  {\bibfnamefont {N.}~\bibnamefont {Wschebor}},\ }\bibfield  {title} {\bibinfo
  {title} {The nonperturbative functional renormalization group and its
  applications},\ }\href
  {https://doi.org/https://doi.org/10.1016/j.physrep.2021.01.001} {\bibfield
  {journal} {\bibinfo  {journal} {Physics Reports}\ }\textbf {\bibinfo {volume}
  {910}},\ \bibinfo {pages} {1} (\bibinfo {year} {2021})}\BibitemShut {NoStop}%
\bibitem [{\citenamefont {Dupuis}\ and\ \citenamefont
  {Daviet}(2020)}]{dupuis2020bose}%
  \BibitemOpen
  \bibfield  {author} {\bibinfo {author} {\bibfnamefont {N.}~\bibnamefont
  {Dupuis}}\ and\ \bibinfo {author} {\bibfnamefont {R.}~\bibnamefont
  {Daviet}},\ }\bibfield  {title} {\bibinfo {title} {Bose-glass phase of a
  one-dimensional disordered bose fluid: Metastable states, quantum tunneling,
  and droplets},\ }\href {https://doi.org/10.1103/PhysRevE.101.042139}
  {\bibfield  {journal} {\bibinfo  {journal} {Phys. Rev. E}\ }\textbf {\bibinfo
  {volume} {101}},\ \bibinfo {pages} {042139} (\bibinfo {year}
  {2020})}\BibitemShut {NoStop}%
\bibitem [{\citenamefont {Kardar}\ \emph {et~al.}(1986)\citenamefont {Kardar},
  \citenamefont {Parisi},\ and\ \citenamefont {Zhang}}]{kardar1986dynamic}%
  \BibitemOpen
  \bibfield  {author} {\bibinfo {author} {\bibfnamefont {M.}~\bibnamefont
  {Kardar}}, \bibinfo {author} {\bibfnamefont {G.}~\bibnamefont {Parisi}},\
  and\ \bibinfo {author} {\bibfnamefont {Y.-C.}\ \bibnamefont {Zhang}},\
  }\bibfield  {title} {\bibinfo {title} {Dynamic scaling of growing
  interfaces},\ }\href {https://doi.org/10.1103/PhysRevLett.56.889} {\bibfield
  {journal} {\bibinfo  {journal} {Phys. Rev. Lett.}\ }\textbf {\bibinfo
  {volume} {56}},\ \bibinfo {pages} {889} (\bibinfo {year} {1986})}\BibitemShut
  {NoStop}%
\bibitem [{\citenamefont {Toner}\ and\ \citenamefont
  {Tu}(1995)}]{toner1995long}%
  \BibitemOpen
  \bibfield  {author} {\bibinfo {author} {\bibfnamefont {J.}~\bibnamefont
  {Toner}}\ and\ \bibinfo {author} {\bibfnamefont {Y.}~\bibnamefont {Tu}},\
  }\bibfield  {title} {\bibinfo {title} {Long-range order in a two-dimensional
  dynamical $\mathrm{XY}$ model: How birds fly together},\ }\href
  {https://doi.org/10.1103/PhysRevLett.75.4326} {\bibfield  {journal} {\bibinfo
   {journal} {Phys. Rev. Lett.}\ }\textbf {\bibinfo {volume} {75}},\ \bibinfo
  {pages} {4326} (\bibinfo {year} {1995})}\BibitemShut {NoStop}%
\bibitem [{\citenamefont {Toner}\ and\ \citenamefont
  {Tu}(1998)}]{toner1998flocks}%
  \BibitemOpen
  \bibfield  {author} {\bibinfo {author} {\bibfnamefont {J.}~\bibnamefont
  {Toner}}\ and\ \bibinfo {author} {\bibfnamefont {Y.}~\bibnamefont {Tu}},\
  }\bibfield  {title} {\bibinfo {title} {Flocks, herds, and schools: A
  quantitative theory of flocking},\ }\href
  {https://doi.org/10.1103/PhysRevE.58.4828} {\bibfield  {journal} {\bibinfo
  {journal} {Phys. Rev. E}\ }\textbf {\bibinfo {volume} {58}},\ \bibinfo
  {pages} {4828} (\bibinfo {year} {1998})}\BibitemShut {NoStop}%
\bibitem [{\citenamefont {Yu}\ \emph {et~al.}(2021)\citenamefont {Yu},
  \citenamefont {Zhang},\ and\ \citenamefont {Tu}}]{yu2021inverse}%
  \BibitemOpen
  \bibfield  {author} {\bibinfo {author} {\bibfnamefont {Q.}~\bibnamefont
  {Yu}}, \bibinfo {author} {\bibfnamefont {D.}~\bibnamefont {Zhang}},\ and\
  \bibinfo {author} {\bibfnamefont {Y.}~\bibnamefont {Tu}},\ }\bibfield
  {title} {\bibinfo {title} {Inverse power law scaling of energy dissipation
  rate in nonequilibrium reaction networks},\ }\href
  {https://doi.org/10.1103/PhysRevLett.126.080601} {\bibfield  {journal}
  {\bibinfo  {journal} {Phys. Rev. Lett.}\ }\textbf {\bibinfo {volume} {126}},\
  \bibinfo {pages} {080601} (\bibinfo {year} {2021})}\BibitemShut {NoStop}%
\bibitem [{\citenamefont {Stephanov}(2004)}]{stephanov2005qcd}%
  \BibitemOpen
  \bibfield  {author} {\bibinfo {author} {\bibfnamefont {M.}~\bibnamefont
  {Stephanov}},\ }\bibfield  {title} {\bibinfo {title} {{QCD Phase Diagram and
  the Critical Point}},\ }\href {https://doi.org/10.1143/PTPS.153.139}
  {\bibfield  {journal} {\bibinfo  {journal} {Progress of Theoretical Physics
  Supplement}\ }\textbf {\bibinfo {volume} {153}},\ \bibinfo {pages} {139}
  (\bibinfo {year} {2004})}\BibitemShut {NoStop}%
\bibitem [{\citenamefont {Murdock}(2003)}]{murdock2003normal}%
  \BibitemOpen
  \bibfield  {author} {\bibinfo {author} {\bibfnamefont {J.~A.}\ \bibnamefont
  {Murdock}},\ }\href@noop {} {\emph {\bibinfo {title} {Normal forms and
  unfoldings for local dynamical systems}}}\ (\bibinfo  {publisher}
  {Springer},\ \bibinfo {year} {2003})\BibitemShut {NoStop}%
\bibitem [{\citenamefont {Raju}\ \emph {et~al.}(2019)\citenamefont {Raju},
  \citenamefont {Clement}, \citenamefont {Hayden}, \citenamefont {Kent-Dobias},
  \citenamefont {Liarte}, \citenamefont {Rocklin},\ and\ \citenamefont
  {Sethna}}]{raju2019normal}%
  \BibitemOpen
  \bibfield  {author} {\bibinfo {author} {\bibfnamefont {A.}~\bibnamefont
  {Raju}}, \bibinfo {author} {\bibfnamefont {C.~B.}\ \bibnamefont {Clement}},
  \bibinfo {author} {\bibfnamefont {L.~X.}\ \bibnamefont {Hayden}}, \bibinfo
  {author} {\bibfnamefont {J.~P.}\ \bibnamefont {Kent-Dobias}}, \bibinfo
  {author} {\bibfnamefont {D.~B.}\ \bibnamefont {Liarte}}, \bibinfo {author}
  {\bibfnamefont {D.~Z.}\ \bibnamefont {Rocklin}},\ and\ \bibinfo {author}
  {\bibfnamefont {J.~P.}\ \bibnamefont {Sethna}},\ }\bibfield  {title}
  {\bibinfo {title} {Normal form for renormalization groups},\ }\href
  {https://doi.org/10.1103/PhysRevX.9.021014} {\bibfield  {journal} {\bibinfo
  {journal} {Phys. Rev. X}\ }\textbf {\bibinfo {volume} {9}},\ \bibinfo {pages}
  {021014} (\bibinfo {year} {2019})}\BibitemShut {NoStop}%
\bibitem [{\citenamefont {Sethna}\ \emph {et~al.}(2023)\citenamefont {Sethna},
  \citenamefont {Hathcock}, \citenamefont {Kent-Dobias},\ and\ \citenamefont
  {Raju}}]{sethna2023normal}%
  \BibitemOpen
  \bibfield  {author} {\bibinfo {author} {\bibfnamefont {J.~P.}\ \bibnamefont
  {Sethna}}, \bibinfo {author} {\bibfnamefont {D.}~\bibnamefont {Hathcock}},
  \bibinfo {author} {\bibfnamefont {J.}~\bibnamefont {Kent-Dobias}},\ and\
  \bibinfo {author} {\bibfnamefont {A.}~\bibnamefont {Raju}},\ }\bibfield
  {title} {\bibinfo {title} {Normal forms, universal scaling functions, and
  extending the validity of the {RG}},\ }\href@noop {} {\bibfield  {journal}
  {\bibinfo  {journal} {arXiv preprint arXiv:2304.00105}\ } (\bibinfo {year}
  {2023})}\BibitemShut {NoStop}%
\bibitem [{\citenamefont {Onsager}(1944)}]{onsager1944crystal}%
  \BibitemOpen
  \bibfield  {author} {\bibinfo {author} {\bibfnamefont {L.}~\bibnamefont
  {Onsager}},\ }\bibfield  {title} {\bibinfo {title} {Crystal statistics. {I}.
  a two-dimensional model with an order-disorder transition},\ }\href
  {https://doi.org/10.1103/PhysRev.65.117} {\bibfield  {journal} {\bibinfo
  {journal} {Phys. Rev.}\ }\textbf {\bibinfo {volume} {65}},\ \bibinfo {pages}
  {117} (\bibinfo {year} {1944})}\BibitemShut {NoStop}%
\bibitem [{Note1()}]{Note1}%
  \BibitemOpen
  \bibinfo {note} {Corrections to scaling due to irrelevant perturbations in
  the 2D Ising model have integer exponents and are indistinguishable from
  analytic corrections. These are known as \protect \emph {redundant}
  variables; there exists a suitable RG with the zero-field 2D Ising model as a
  fixed point \cite {raju2018reexamining}.}\BibitemShut {Stop}%
\bibitem [{Note2()}]{Note2}%
  \BibitemOpen
  \bibinfo {note} {In the case of the 2D Ising model the quadratic temperature
  term in the free energy $At^2$ is non-removable due to a resonance: an
  eigenvalue of the of the RG flow is an integral linear combination of other
  eigenvalues, $\lambda _0 = \DOTSB \sum@ \slimits@ _i m_i \lambda _i$, with
  $m_i \in \protect \mathbb {N}$ \cite {murdock2003normal, raju2019normal}.
  Specifically, temperature $\lambda _t=1$ and the free energy $\lambda _f = 2
  = 2 \lambda _t$ are resonant.}\BibitemShut {Stop}%
\bibitem [{SM()}]{SM}%
  \BibitemOpen
  \href@noop {} {}\bibinfo {note} {See Supplemental Material at [url] for
  expansions of the nonlinear renormalization group flows, radius of
  convergence analysis, complex-temperature free-energy diagrams, and details
  of the 3D Ising fitting procedure.}\BibitemShut {Stop}%
\bibitem [{\citenamefont {Orrick}\ \emph {et~al.}(2001)\citenamefont {Orrick},
  \citenamefont {Nickel}, \citenamefont {Guttmann},\ and\ \citenamefont
  {Perk}}]{orrick2001critical}%
  \BibitemOpen
  \bibfield  {author} {\bibinfo {author} {\bibfnamefont {W.~P.}\ \bibnamefont
  {Orrick}}, \bibinfo {author} {\bibfnamefont {B.~G.}\ \bibnamefont {Nickel}},
  \bibinfo {author} {\bibfnamefont {A.~J.}\ \bibnamefont {Guttmann}},\ and\
  \bibinfo {author} {\bibfnamefont {J.~H.~H.}\ \bibnamefont {Perk}},\
  }\bibfield  {title} {\bibinfo {title} {Critical behavior of the
  two-dimensional {I}sing susceptibility},\ }\href
  {https://doi.org/10.1103/PhysRevLett.86.4120} {\bibfield  {journal} {\bibinfo
   {journal} {Phys. Rev. Lett.}\ }\textbf {\bibinfo {volume} {86}},\ \bibinfo
  {pages} {4120} (\bibinfo {year} {2001})}\BibitemShut {NoStop}%
\bibitem [{\citenamefont {Caselle}\ \emph {et~al.}(2002)\citenamefont
  {Caselle}, \citenamefont {Hasenbusch}, \citenamefont {Pelissetto},\ and\
  \citenamefont {Vicari}}]{caselle2002irrelevant}%
  \BibitemOpen
  \bibfield  {author} {\bibinfo {author} {\bibfnamefont {M.}~\bibnamefont
  {Caselle}}, \bibinfo {author} {\bibfnamefont {M.}~\bibnamefont {Hasenbusch}},
  \bibinfo {author} {\bibfnamefont {A.}~\bibnamefont {Pelissetto}},\ and\
  \bibinfo {author} {\bibfnamefont {E.}~\bibnamefont {Vicari}},\ }\bibfield
  {title} {\bibinfo {title} {Irrelevant operators in the two-dimensional
  {I}sing model},\ }\href {https://doi.org/10.1088/0305-4470/35/23/305}
  {\bibfield  {journal} {\bibinfo  {journal} {Journal of Physics A:
  Mathematical and General}\ }\textbf {\bibinfo {volume} {35}},\ \bibinfo
  {pages} {4861} (\bibinfo {year} {2002})}\BibitemShut {NoStop}%
\bibitem [{\citenamefont {Chan}\ \emph {et~al.}(2011)\citenamefont {Chan},
  \citenamefont {Guttmann}, \citenamefont {Nickel},\ and\ \citenamefont
  {Perk}}]{chan2011ising}%
  \BibitemOpen
  \bibfield  {author} {\bibinfo {author} {\bibfnamefont {Y.}~\bibnamefont
  {Chan}}, \bibinfo {author} {\bibfnamefont {A.~J.}\ \bibnamefont {Guttmann}},
  \bibinfo {author} {\bibfnamefont {B.~G.}\ \bibnamefont {Nickel}},\ and\
  \bibinfo {author} {\bibfnamefont {J.~H.~H.}\ \bibnamefont {Perk}},\
  }\bibfield  {title} {\bibinfo {title} {The {I}sing susceptibility scaling
  function},\ }\href {https://doi.org/10.1007/s10955-011-0212-0} {\bibfield
  {journal} {\bibinfo  {journal} {Journal of Statistical Physics}\ }\textbf
  {\bibinfo {volume} {145}},\ \bibinfo {pages} {549} (\bibinfo {year}
  {2011})}\BibitemShut {NoStop}%
\bibitem [{\citenamefont {Fisher}(1964)}]{fisher1964nature}%
  \BibitemOpen
  \bibfield  {author} {\bibinfo {author} {\bibfnamefont {M.~E.}\ \bibnamefont
  {Fisher}},\ }\bibfield  {title} {\bibinfo {title} {The nature of critical
  points},\ }in\ \href@noop {} {\emph {\bibinfo {booktitle} {Boulder Lectures
  in Theoretical Physics: Statistical Physics, Weak Interactions, Field
  Theory}}},\ Vol.~\bibinfo {volume} {7c}\ (\bibinfo  {publisher} {University
  of Colorado Press},\ \bibinfo {year} {1964})\ pp.\ \bibinfo {pages}
  {1--159}\BibitemShut {NoStop}%
\bibitem [{\citenamefont {Dolan}\ \emph {et~al.}(2001)\citenamefont {Dolan},
  \citenamefont {Janke}, \citenamefont {Johnston},\ and\ \citenamefont
  {Stathakopoulos}}]{dolan2001thin}%
  \BibitemOpen
  \bibfield  {author} {\bibinfo {author} {\bibfnamefont {B.~P.}\ \bibnamefont
  {Dolan}}, \bibinfo {author} {\bibfnamefont {W.}~\bibnamefont {Janke}},
  \bibinfo {author} {\bibfnamefont {D.~A.}\ \bibnamefont {Johnston}},\ and\
  \bibinfo {author} {\bibfnamefont {M.}~\bibnamefont {Stathakopoulos}},\
  }\bibfield  {title} {\bibinfo {title} {Thin {F}isher zeros},\ }\href
  {https://doi.org/10.1088/0305-4470/34/32/301} {\bibfield  {journal} {\bibinfo
   {journal} {Journal of Physics A: Mathematical and General}\ }\textbf
  {\bibinfo {volume} {34}},\ \bibinfo {pages} {6211} (\bibinfo {year}
  {2001})}\BibitemShut {NoStop}%
\bibitem [{\citenamefont {Denbleyker}\ \emph {et~al.}(2010)\citenamefont
  {Denbleyker}, \citenamefont {Du}, \citenamefont {Liu}, \citenamefont
  {Meurice},\ and\ \citenamefont {Zou}}]{denbleyker2010fisher}%
  \BibitemOpen
  \bibfield  {author} {\bibinfo {author} {\bibfnamefont {A.}~\bibnamefont
  {Denbleyker}}, \bibinfo {author} {\bibfnamefont {D.}~\bibnamefont {Du}},
  \bibinfo {author} {\bibfnamefont {Y.}~\bibnamefont {Liu}}, \bibinfo {author}
  {\bibfnamefont {Y.}~\bibnamefont {Meurice}},\ and\ \bibinfo {author}
  {\bibfnamefont {H.}~\bibnamefont {Zou}},\ }\bibfield  {title} {\bibinfo
  {title} {Fisher's zeros as the boundary of renormalization group flows in
  complex coupling spaces},\ }\href
  {https://doi.org/10.1103/PhysRevLett.104.251601} {\bibfield  {journal}
  {\bibinfo  {journal} {Phys. Rev. Lett.}\ }\textbf {\bibinfo {volume} {104}},\
  \bibinfo {pages} {251601} (\bibinfo {year} {2010})}\BibitemShut {NoStop}%
\bibitem [{\citenamefont {Campbell}\ \emph {et~al.}(2007)\citenamefont
  {Campbell}, \citenamefont {Hukushima},\ and\ \citenamefont
  {Takayama}}]{campbell2007extended}%
  \BibitemOpen
  \bibfield  {author} {\bibinfo {author} {\bibfnamefont {I.~A.}\ \bibnamefont
  {Campbell}}, \bibinfo {author} {\bibfnamefont {K.}~\bibnamefont
  {Hukushima}},\ and\ \bibinfo {author} {\bibfnamefont {H.}~\bibnamefont
  {Takayama}},\ }\bibfield  {title} {\bibinfo {title} {Extended scaling for
  ferromagnets},\ }\href {https://doi.org/10.1103/PhysRevB.76.134421}
  {\bibfield  {journal} {\bibinfo  {journal} {Phys. Rev. B}\ }\textbf {\bibinfo
  {volume} {76}},\ \bibinfo {pages} {134421} (\bibinfo {year}
  {2007})}\BibitemShut {NoStop}%
\bibitem [{\citenamefont {Campbell}\ and\ \citenamefont
  {Butera}(2008)}]{campbell2008extended}%
  \BibitemOpen
  \bibfield  {author} {\bibinfo {author} {\bibfnamefont {I.~A.}\ \bibnamefont
  {Campbell}}\ and\ \bibinfo {author} {\bibfnamefont {P.}~\bibnamefont
  {Butera}},\ }\bibfield  {title} {\bibinfo {title} {Extended scaling for the
  high-dimension and square-lattice ising ferromagnets},\ }\href
  {https://doi.org/10.1103/PhysRevB.78.024435} {\bibfield  {journal} {\bibinfo
  {journal} {Phys. Rev. B}\ }\textbf {\bibinfo {volume} {78}},\ \bibinfo
  {pages} {024435} (\bibinfo {year} {2008})}\BibitemShut {NoStop}%
\bibitem [{\citenamefont {Katzgraber}\ \emph {et~al.}(2008)\citenamefont
  {Katzgraber}, \citenamefont {Campbell},\ and\ \citenamefont
  {Hartmann}}]{campbell2008extendedZeroTemp}%
  \BibitemOpen
  \bibfield  {author} {\bibinfo {author} {\bibfnamefont {H.~G.}\ \bibnamefont
  {Katzgraber}}, \bibinfo {author} {\bibfnamefont {I.~A.}\ \bibnamefont
  {Campbell}},\ and\ \bibinfo {author} {\bibfnamefont {A.~K.}\ \bibnamefont
  {Hartmann}},\ }\bibfield  {title} {\bibinfo {title} {Extended scaling for
  ferromagnetic ising models with zero-temperature transitions},\ }\href
  {https://doi.org/10.1103/PhysRevB.78.184409} {\bibfield  {journal} {\bibinfo
  {journal} {Phys. Rev. B}\ }\textbf {\bibinfo {volume} {78}},\ \bibinfo
  {pages} {184409} (\bibinfo {year} {2008})}\BibitemShut {NoStop}%
\bibitem [{\citenamefont {Campbell}\ and\ \citenamefont
  {Lundow}(2011)}]{campbell2011extended}%
  \BibitemOpen
  \bibfield  {author} {\bibinfo {author} {\bibfnamefont {I.~A.}\ \bibnamefont
  {Campbell}}\ and\ \bibinfo {author} {\bibfnamefont {P.~H.}\ \bibnamefont
  {Lundow}},\ }\bibfield  {title} {\bibinfo {title} {Extended scaling analysis
  of the $s=\frac{1}{2}$ ising ferromagnet on the simple cubic lattice},\
  }\href {https://doi.org/10.1103/PhysRevB.83.014411} {\bibfield  {journal}
  {\bibinfo  {journal} {Phys. Rev. B}\ }\textbf {\bibinfo {volume} {83}},\
  \bibinfo {pages} {014411} (\bibinfo {year} {2011})}\BibitemShut {NoStop}%
\bibitem [{\citenamefont {Kos}\ \emph {et~al.}(2016)\citenamefont {Kos},
  \citenamefont {Poland}, \citenamefont {Simmons-Duffin},\ and\ \citenamefont
  {Vichi}}]{kos2016precision}%
  \BibitemOpen
  \bibfield  {author} {\bibinfo {author} {\bibfnamefont {F.}~\bibnamefont
  {Kos}}, \bibinfo {author} {\bibfnamefont {D.}~\bibnamefont {Poland}},
  \bibinfo {author} {\bibfnamefont {D.}~\bibnamefont {Simmons-Duffin}},\ and\
  \bibinfo {author} {\bibfnamefont {A.}~\bibnamefont {Vichi}},\ }\bibfield
  {title} {\bibinfo {title} {Precision islands in the {I}sing and {O(N)}
  models},\ }\href {https://doi.org/10.1007/JHEP08(2016)036} {\bibfield
  {journal} {\bibinfo  {journal} {Journal of High Energy Physics}\ }\textbf
  {\bibinfo {volume} {2016}},\ \bibinfo {pages} {36} (\bibinfo {year}
  {2016})}\BibitemShut {NoStop}%
\bibitem [{\citenamefont {Komargodski}\ and\ \citenamefont
  {Simmons-Duffin}(2017)}]{komargodski2017random}%
  \BibitemOpen
  \bibfield  {author} {\bibinfo {author} {\bibfnamefont {Z.}~\bibnamefont
  {Komargodski}}\ and\ \bibinfo {author} {\bibfnamefont {D.}~\bibnamefont
  {Simmons-Duffin}},\ }\bibfield  {title} {\bibinfo {title} {The random-bond
  {I}sing model in 2.01 and 3 dimensions},\ }\href
  {https://doi.org/10.1088/1751-8121/aa6087} {\bibfield  {journal} {\bibinfo
  {journal} {Journal of Physics A: Mathematical and Theoretical}\ }\textbf
  {\bibinfo {volume} {50}},\ \bibinfo {pages} {154001} (\bibinfo {year}
  {2017})}\BibitemShut {NoStop}%
\bibitem [{\citenamefont {Hasenbusch}(2010)}]{hasenbusch2010universal}%
  \BibitemOpen
  \bibfield  {author} {\bibinfo {author} {\bibfnamefont {M.}~\bibnamefont
  {Hasenbusch}},\ }\bibfield  {title} {\bibinfo {title} {Universal amplitude
  ratios in the three-dimensional {I}sing universality class},\ }\href
  {https://doi.org/10.1103/PhysRevB.82.174434} {\bibfield  {journal} {\bibinfo
  {journal} {Phys. Rev. B}\ }\textbf {\bibinfo {volume} {82}},\ \bibinfo
  {pages} {174434} (\bibinfo {year} {2010})}\BibitemShut {NoStop}%
\bibitem [{\citenamefont {Guttmann}\ and\ \citenamefont
  {Enting}(1993)}]{guttmann1993series}%
  \BibitemOpen
  \bibfield  {author} {\bibinfo {author} {\bibfnamefont {A.~J.}\ \bibnamefont
  {Guttmann}}\ and\ \bibinfo {author} {\bibfnamefont {I.~G.}\ \bibnamefont
  {Enting}},\ }\bibfield  {title} {\bibinfo {title} {Series studies of the
  {P}otts model. {I.} the simple cubic {I}sing model},\ }\href
  {https://doi.org/10.1088/0305-4470/26/4/010} {\bibfield  {journal} {\bibinfo
  {journal} {Journal of Physics A: Mathematical and General}\ }\textbf
  {\bibinfo {volume} {26}},\ \bibinfo {pages} {807} (\bibinfo {year}
  {1993})}\BibitemShut {NoStop}%
\bibitem [{\citenamefont {Guttmann}\ and\ \citenamefont
  {Enting}(1994)}]{guttmann1994high}%
  \BibitemOpen
  \bibfield  {author} {\bibinfo {author} {\bibfnamefont {A.~J.}\ \bibnamefont
  {Guttmann}}\ and\ \bibinfo {author} {\bibfnamefont {I.~G.}\ \bibnamefont
  {Enting}},\ }\bibfield  {title} {\bibinfo {title} {The high-temperature
  specific heat exponent of the 3d {I}sing model},\ }\href
  {https://doi.org/10.1088/0305-4470/27/24/012} {\bibfield  {journal} {\bibinfo
   {journal} {Journal of Physics A: Mathematical and General}\ }\textbf
  {\bibinfo {volume} {27}},\ \bibinfo {pages} {8007} (\bibinfo {year}
  {1994})}\BibitemShut {NoStop}%
\bibitem [{\citenamefont {Bhanot}\ \emph {et~al.}(1994)\citenamefont {Bhanot},
  \citenamefont {Creutz}, \citenamefont {Horvath}, \citenamefont {Lacki},\ and\
  \citenamefont {Weckel}}]{bhanot1994series}%
  \BibitemOpen
  \bibfield  {author} {\bibinfo {author} {\bibfnamefont {G.}~\bibnamefont
  {Bhanot}}, \bibinfo {author} {\bibfnamefont {M.}~\bibnamefont {Creutz}},
  \bibinfo {author} {\bibfnamefont {I.}~\bibnamefont {Horvath}}, \bibinfo
  {author} {\bibfnamefont {J.}~\bibnamefont {Lacki}},\ and\ \bibinfo {author}
  {\bibfnamefont {J.}~\bibnamefont {Weckel}},\ }\bibfield  {title} {\bibinfo
  {title} {Series expansions without diagrams},\ }\href
  {https://doi.org/10.1103/PhysRevE.49.2445} {\bibfield  {journal} {\bibinfo
  {journal} {Phys. Rev. E}\ }\textbf {\bibinfo {volume} {49}},\ \bibinfo
  {pages} {2445} (\bibinfo {year} {1994})}\BibitemShut {NoStop}%
\bibitem [{\citenamefont {Essam}\ and\ \citenamefont
  {Fisher}(1963)}]{essam1963pade}%
  \BibitemOpen
  \bibfield  {author} {\bibinfo {author} {\bibfnamefont {J.~W.}\ \bibnamefont
  {Essam}}\ and\ \bibinfo {author} {\bibfnamefont {M.~E.}\ \bibnamefont
  {Fisher}},\ }\bibfield  {title} {\bibinfo {title} {{Padé Approximant Studies
  of the Lattice Gas and Ising Ferromagnet below the Critical Point}},\ }\href
  {https://doi.org/10.1063/1.1733766} {\bibfield  {journal} {\bibinfo
  {journal} {The Journal of Chemical Physics}\ }\textbf {\bibinfo {volume}
  {38}},\ \bibinfo {pages} {802} (\bibinfo {year} {1963})}\BibitemShut
  {NoStop}%
\bibitem [{\citenamefont {Talapov}\ and\ \citenamefont
  {Blöte}(1996)}]{talapov1996magnetization}%
  \BibitemOpen
  \bibfield  {author} {\bibinfo {author} {\bibfnamefont {A.~L.}\ \bibnamefont
  {Talapov}}\ and\ \bibinfo {author} {\bibfnamefont {H.~W.~J.}\ \bibnamefont
  {Blöte}},\ }\bibfield  {title} {\bibinfo {title} {The magnetization of the
  3d ising model},\ }\href {https://doi.org/10.1088/0305-4470/29/17/042}
  {\bibfield  {journal} {\bibinfo  {journal} {Journal of Physics A:
  Mathematical and General}\ }\textbf {\bibinfo {volume} {29}},\ \bibinfo
  {pages} {5727} (\bibinfo {year} {1996})}\BibitemShut {NoStop}%
\bibitem [{\citenamefont {Hasenbusch}(2012)}]{hasenbusch2012thermodynamic}%
  \BibitemOpen
  \bibfield  {author} {\bibinfo {author} {\bibfnamefont {M.}~\bibnamefont
  {Hasenbusch}},\ }\bibfield  {title} {\bibinfo {title} {Thermodynamic casimir
  effect: Universality and corrections to scaling},\ }\href
  {https://doi.org/10.1103/PhysRevB.85.174421} {\bibfield  {journal} {\bibinfo
  {journal} {Phys. Rev. B}\ }\textbf {\bibinfo {volume} {85}},\ \bibinfo
  {pages} {174421} (\bibinfo {year} {2012})}\BibitemShut {NoStop}%
\bibitem [{\citenamefont {Raju}(2018)}]{raju2018aspects}%
  \BibitemOpen
  \bibfield  {author} {\bibinfo {author} {\bibfnamefont {A.}~\bibnamefont
  {Raju}},\ }\emph {\bibinfo {title} {Aspects of the Renormalization Group}},\
  \href@noop {} {Ph.D. thesis},\ \bibinfo  {school} {Cornell University}
  (\bibinfo {year} {2018})\BibitemShut {NoStop}%
\bibitem [{\citenamefont {Clement}(2019)}]{clement2019respect}%
  \BibitemOpen
  \bibfield  {author} {\bibinfo {author} {\bibfnamefont {C.}~\bibnamefont
  {Clement}},\ }\href@noop {} {\emph {\bibinfo {title} {Respect Your Data:
  Topics in Inference and Modeling in Physics}}}\ (\bibinfo  {publisher}
  {Cornell University},\ \bibinfo {year} {2019})\BibitemShut {NoStop}%
\bibitem [{\citenamefont {Sykes}\ \emph {et~al.}(1965)\citenamefont {Sykes},
  \citenamefont {Essam},\ and\ \citenamefont {Gaunt}}]{sykes1965derivation}%
  \BibitemOpen
  \bibfield  {author} {\bibinfo {author} {\bibfnamefont {M.~F.}\ \bibnamefont
  {Sykes}}, \bibinfo {author} {\bibfnamefont {J.~W.}\ \bibnamefont {Essam}},\
  and\ \bibinfo {author} {\bibfnamefont {D.~S.}\ \bibnamefont {Gaunt}},\
  }\bibfield  {title} {\bibinfo {title} {{Derivation of Low‐Temperature
  Expansions for the {I}sing Model of a Ferromagnet and an Antiferromagnet}},\
  }\href {https://doi.org/10.1063/1.1704279} {\bibfield  {journal} {\bibinfo
  {journal} {Journal of Mathematical Physics}\ }\textbf {\bibinfo {volume}
  {6}},\ \bibinfo {pages} {283} (\bibinfo {year} {1965})}\BibitemShut {NoStop}%
\bibitem [{\citenamefont {Katsura}\ \emph {et~al.}(1977)\citenamefont
  {Katsura}, \citenamefont {Yazaki},\ and\ \citenamefont
  {Takaishi}}]{katsura1977high}%
  \BibitemOpen
  \bibfield  {author} {\bibinfo {author} {\bibfnamefont {S.}~\bibnamefont
  {Katsura}}, \bibinfo {author} {\bibfnamefont {N.}~\bibnamefont {Yazaki}},\
  and\ \bibinfo {author} {\bibfnamefont {M.}~\bibnamefont {Takaishi}},\
  }\bibfield  {title} {\bibinfo {title} {The high temperature – low field
  expansion of the {I}sing model from the low temperature – high field
  expansion},\ }\href {https://doi.org/10.1139/p77-210} {\bibfield  {journal}
  {\bibinfo  {journal} {Canadian Journal of Physics}\ }\textbf {\bibinfo
  {volume} {55}},\ \bibinfo {pages} {1648} (\bibinfo {year}
  {1977})}\BibitemShut {NoStop}%
\bibitem [{\citenamefont {Kent-Dobias}\ and\ \citenamefont
  {Sethna}(2017)}]{kentDobias2017smooth}%
  \BibitemOpen
  \bibfield  {author} {\bibinfo {author} {\bibfnamefont {J.}~\bibnamefont
  {Kent-Dobias}}\ and\ \bibinfo {author} {\bibfnamefont {J.~P.}\ \bibnamefont
  {Sethna}},\ }\bibfield  {title} {\bibinfo {title} {Smooth and global {I}sing
  universal scaling functions},\ }\href@noop {} {\bibfield  {journal} {\bibinfo
   {journal} {arXiv preprint arXiv:1707.03791}\ } (\bibinfo {year}
  {2017})}\BibitemShut {NoStop}%
\bibitem [{\citenamefont {Bors\'anyi}\ \emph {et~al.}(2021)\citenamefont
  {Bors\'anyi}, \citenamefont {Fodor}, \citenamefont {Guenther}, \citenamefont
  {Kara}, \citenamefont {Katz}, \citenamefont {Parotto}, \citenamefont
  {P\'asztor}, \citenamefont {Ratti},\ and\ \citenamefont
  {Szab\'o}}]{borsanyi2021lattice}%
  \BibitemOpen
  \bibfield  {author} {\bibinfo {author} {\bibfnamefont {S.}~\bibnamefont
  {Bors\'anyi}}, \bibinfo {author} {\bibfnamefont {Z.}~\bibnamefont {Fodor}},
  \bibinfo {author} {\bibfnamefont {J.~N.}\ \bibnamefont {Guenther}}, \bibinfo
  {author} {\bibfnamefont {R.}~\bibnamefont {Kara}}, \bibinfo {author}
  {\bibfnamefont {S.~D.}\ \bibnamefont {Katz}}, \bibinfo {author}
  {\bibfnamefont {P.}~\bibnamefont {Parotto}}, \bibinfo {author} {\bibfnamefont
  {A.}~\bibnamefont {P\'asztor}}, \bibinfo {author} {\bibfnamefont
  {C.}~\bibnamefont {Ratti}},\ and\ \bibinfo {author} {\bibfnamefont {K.~K.}\
  \bibnamefont {Szab\'o}},\ }\bibfield  {title} {\bibinfo {title} {Lattice
  {QCD} equation of state at finite chemical potential from an alternative
  expansion scheme},\ }\href {https://doi.org/10.1103/PhysRevLett.126.232001}
  {\bibfield  {journal} {\bibinfo  {journal} {Phys. Rev. Lett.}\ }\textbf
  {\bibinfo {volume} {126}},\ \bibinfo {pages} {232001} (\bibinfo {year}
  {2021})}\BibitemShut {NoStop}%
\bibitem [{\citenamefont {Machta}\ \emph {et~al.}(2011)\citenamefont {Machta},
  \citenamefont {Papanikolaou}, \citenamefont {Sethna},\ and\ \citenamefont
  {Veatch}}]{machta2011minimal}%
  \BibitemOpen
  \bibfield  {author} {\bibinfo {author} {\bibfnamefont {B.}~\bibnamefont
  {Machta}}, \bibinfo {author} {\bibfnamefont {S.}~\bibnamefont
  {Papanikolaou}}, \bibinfo {author} {\bibfnamefont {J.}~\bibnamefont
  {Sethna}},\ and\ \bibinfo {author} {\bibfnamefont {S.}~\bibnamefont
  {Veatch}},\ }\bibfield  {title} {\bibinfo {title} {Minimal model of plasma
  membrane heterogeneity requires coupling cortical actin to criticality},\
  }\href {https://doi.org/https://doi.org/10.1016/j.bpj.2011.02.029} {\bibfield
   {journal} {\bibinfo  {journal} {Biophysical Journal}\ }\textbf {\bibinfo
  {volume} {100}},\ \bibinfo {pages} {1668} (\bibinfo {year}
  {2011})}\BibitemShut {NoStop}%
\bibitem [{\citenamefont {Machta}\ \emph {et~al.}(2012)\citenamefont {Machta},
  \citenamefont {Veatch},\ and\ \citenamefont {Sethna}}]{machta2012critical}%
  \BibitemOpen
  \bibfield  {author} {\bibinfo {author} {\bibfnamefont {B.~B.}\ \bibnamefont
  {Machta}}, \bibinfo {author} {\bibfnamefont {S.~L.}\ \bibnamefont {Veatch}},\
  and\ \bibinfo {author} {\bibfnamefont {J.~P.}\ \bibnamefont {Sethna}},\
  }\bibfield  {title} {\bibinfo {title} {Critical casimir forces in cellular
  membranes},\ }\href {https://doi.org/10.1103/PhysRevLett.109.138101}
  {\bibfield  {journal} {\bibinfo  {journal} {Phys. Rev. Lett.}\ }\textbf
  {\bibinfo {volume} {109}},\ \bibinfo {pages} {138101} (\bibinfo {year}
  {2012})}\BibitemShut {NoStop}%
\bibitem [{\citenamefont {Shankar}(1994)}]{shankar1994renormalization}%
  \BibitemOpen
  \bibfield  {author} {\bibinfo {author} {\bibfnamefont {R.}~\bibnamefont
  {Shankar}},\ }\bibfield  {title} {\bibinfo {title} {Renormalization-group
  approach to interacting fermions},\ }\href
  {https://doi.org/10.1103/RevModPhys.66.129} {\bibfield  {journal} {\bibinfo
  {journal} {Rev. Mod. Phys.}\ }\textbf {\bibinfo {volume} {66}},\ \bibinfo
  {pages} {129} (\bibinfo {year} {1994})}\BibitemShut {NoStop}%
\bibitem [{\citenamefont {Abanov}\ and\ \citenamefont
  {Chubukov}(2000)}]{abanov2001spin}%
  \BibitemOpen
  \bibfield  {author} {\bibinfo {author} {\bibfnamefont {A.}~\bibnamefont
  {Abanov}}\ and\ \bibinfo {author} {\bibfnamefont {A.~V.}\ \bibnamefont
  {Chubukov}},\ }\bibfield  {title} {\bibinfo {title} {Spin-fermion model near
  the quantum critical point: One-loop renormalization group results},\ }\href
  {https://doi.org/10.1103/PhysRevLett.84.5608} {\bibfield  {journal} {\bibinfo
   {journal} {Phys. Rev. Lett.}\ }\textbf {\bibinfo {volume} {84}},\ \bibinfo
  {pages} {5608} (\bibinfo {year} {2000})}\BibitemShut {NoStop}%
\end{thebibliography}%

\clearpage

\end{document}